\documentclass[12pt]{iopart}
\usepackage{graphicx}
\usepackage{multirow}
\usepackage{hyperref}
\usepackage{iopams}
\usepackage{cite}

\begin{document}

\title[Absolute cross sections for 3d photoionization of Xe$^{q+}$ ions ($1\leq q \leq 5$)]{Absolute cross sections for photoionization of Xe$^\mathbf{q+}$ ions ($\mathbf{1\leq q \leq5}$)  at the 3d ionization threshold}

\author{S.~Schippers$^1$, S.~Ricz$^1$\footnote{permanent address: Institute of Nuclear Research of the Hungarian Academy of Sciences, Debrecen, P.O. Box 51, H-4001, Hungary},
T.~Buhr$^{1,7}$, A.~Borovik Jr.$^1$, J.~Hellhund$^1$, K.~Holste$^1$, K.~Huber$^1$,
H.-J.~Sch\"{a}fer$^1$, D.~Schury$^1$, S.~Klumpp$^2$, K.~Mertens$^2$, M.~Martins$^2$, R.~Flesch$^3$,
G.~Ulrich$^3$, E.~R\"{u}hl$^3$, T.~Jahnke$^4$, J.~Lower$^4$, D.~Metz$^4$, L.~P.~H.~Schmidt$^4$, M.~Sch\"{o}ffler$^4$, J.~B.~Williams$^4$,
L.~Glaser$^5$, F.~Scholz$^5$, J.~Seltmann$^5$, J.~Viefhaus$^5$,
A.~Dorn$^6$, A.~Wolf$^{\,6}$, J.~Ullrich$^7$, and A.~M\"{u}ller$^1$}

\address{$^1$ Institut f\"{u}r Atom- und Molek\"{u}lphysik, Justus-Liebig-Universit\"{a}t Giessen, Leihgesterner Weg 217, 35392 Giessen, Germany}
\address{$^2$ Institut f\"{u}r Experimentalphysik, Universit\"{a}t Hamburg, 22761 Hamburg, Germany}
\address{$^3$ Institut f\"{u}r Chemie und Biochemie, Freie Universit\"{a}t Berlin, 14195 Berlin, Germany}
\address{$^4$ Institut f\"{u}r Kernphysik, Goethe-Universit\"{a}t Frankfurt, 60438 Frankfurt am Main, Germany}
\address{$^5$ FS-PE, DESY, 22607 Hamburg, Germany}
\address{$^6$ Max-Planck-Institut f\"{u}r Kernphysik, Heidelberg, 69117 Heidelberg, Germany}
\address{$^7$ Physikalisch-Technische Bundesanstalt, 38116 Braunschweig, Germany}
\ead{stefan.schippers@physik.uni-giessen.de}

\begin{abstract}

The photon-ion merged-beams technique has been employed at the new \underline{P}hoton-\underline{I}on spectrometer at \underline{PE}TRA\,III  (PIPE) for measuring multiple photoionization of Xe$^{q+}$ ($q$=1--5) ions.  Total ionization cross sections have been obtained on an absolute scale for the dominant ionization reactions of the type $h\nu + \mathrm{Xe}^{q+} \to \mathrm{Xe}^{r+} + (q-r)e^-$ with product charge states $q+2\leq r \leq q+5$. Prominent ionization features are observed in the photon-energy range 650--750 eV, which are associated with excitation or ionization of an inner-shell $3d$ electron. Single-configuration Dirac-Fock calculations agree quantitatively with the experimental cross sections for non-resonant photoabsorption, but fail to reproduce all details of the measured ionization resonance structures.
\end{abstract}

\maketitle

\section{Introduction}

Inner-shell photoionization of heavy atoms with subsequent deexcitation via radiative and Auger cascades leads to the formation of multiply charged ions. The distribution of final charge states depends on the branching ratios for the various deexcitation processes. Comprehensive theoretical calculations for these quantities are cumbersome \cite{Jonauskas2003}, nevertheless, they are required, e.g., for predicting the charge balance in astrophysical plasmas \cite{Kaastra1993} or EUV-light sources \cite{Stamm2004}. Here, we report experimental absolute cross sections for multiple photoionization of Xe$^{q+}$ ions with primary charge states $1 \leq q \leq 5$ which were obtained by employing the photon-ion merged-beams technique at a synchrotron light source. The experimental photon energy range was 500--1200~eV.

The most prominent features of the xenon photoionization cross sections in this energy range are due to inner shell ionization of a $3d$ electron.
In the past, inner-shell $3d$-photoabsorption and $3d$-photoionization of neutral xenon \emph{atoms} has been extensively studied both experimentally \cite{Lukirskii1964,Deslattes1968,Siegbahn1970,Gelius1974,Svensson1976,Yagci1983,Sonntag1984,Becker1987,Becker1989,Saito1992,Arp1999,Karvonen1999,Kivimaeki2000,Sankari2001,Tamenori2002,Jonauskas2003,Matsui2004,Partanen2005,Viefhaus2005,Kochur2009,Suzuki2011} and theoretically \cite{Manson1968,Kennedy1972,Amusia1978,Ohno1982,Zangwill1984,Amusia1985,Kutzner1989,Tong1990,Radojevic2003,Toffoli2003a,Kochur2004,Amusia2007a}. To the best of our knowledge there are no such studies for xenon \emph{ions}, except for the theoretical work of Tong \etal \cite{Tong1990} on $3d\to \varepsilon f$ partial ionization of Xe$^+$ and Xe$^{2+}$. Experimentally, photoionization of ions from the xenon and neighbouring isonuclear sequences has only been studied at lower photon energies, in particular, in the range of the $4d$ ionization threshold \cite{Kjeldsen2002b,Lysaght2005,Bizau2006b,Habibi2009} (and references therein).

The present study extends the work of Saito and Suzuki \cite{Saito1992} who measured yields of multiply charged Xe ions after $3d$ photoionization of neutral xenon atoms. Here we present similar measurements for Xe ions. In addition, the merged-beams technique applied here allows us to put the measured ion yields on an absolute cross section scale. Before presenting and discussing our results we give a comprehensive description of our new photon-ion merged-beams setup and of the experimental procedures.

\section{Experimental setup and experimental procedures}\label{sec:exp}

\begin{figure}
\begin{indented}
\item[]\includegraphics[width=0.8\textwidth]{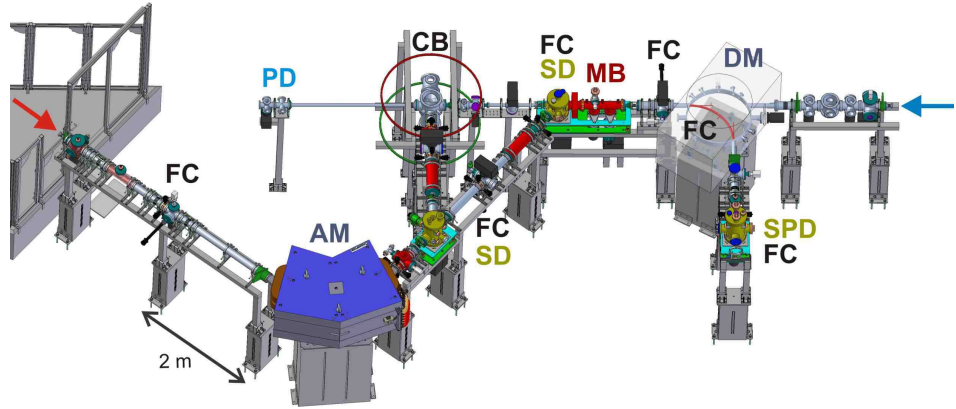}
\end{indented}
\caption{\label{fig:PIPE} (color online) Sketch of the PIPE setup. The photon beam enters the setup from the right (blue arrow) and is parallel to the floor at a nominal height of 2.07~m. It is stopped by a calibrated photodiode (PD) which continuously monitors the absolute photon flux. The ion beam enters from the left (red arrow). It is generated with an ion source that is mounted on a separate platform (not fully shown). The analyzing magnet (AM) provides mass/charge selection of ions for the further ion-beam transport. Spherical deflectors (SD) can be used to direct the ions either into the crossed-beams (CB) interaction point or into the merged-beams (MB) collinear beam overlap region. The demerging magnet (DM) deflects primary and product ions out of the photon-beam axis and directs product ions into the single-particle detector (SPD). The ion current can be measured at various places along the ion beamline by inserting Faraday cups (FC) into the ion beam. One FC is mounted inside the DM such that product ions which are deflected by $90^\circ$ can pass towards the SPD, while ions in different charge states or with different kinetic energies are collected in this FC. The MB is equipped with scanning slits for beam-profile measurements.}
\end{figure}

The well-established photon-ion merged-beams technique \cite{Lyon1986,Wuilleumier1994,Koizumi1995,West2001a,Kjeldsen2006a,Covington2002,Gharaibeh2011a} has been employed at the newly built \underline{P}hoton-\underline{I}on spectrometer at \underline{PE}TRA\,III (PIPE, \fref{fig:PIPE}) \cite{Ricsoka2009a,Schippers2012b}. This experimental setup has been permanently installed as an end station at the "Variable Polarization XUV Beamline" (P04)  of the synchrotron light source PETRA\,III at DESY in Hamburg, Germany. In comparison with similar existing installations at the ALS \cite{Covington2002} in Berkeley, USA, at ASTRID \cite{Kjeldsen2006a} in Aarhus, Denmark, and at SOLEIL \cite{Gharaibeh2011a} in Saint-Auban, France, PIPE is the only  photon-ion merged-beams setup that provides access to photon energies beyond 1000~eV.

\subsection{Photon beamline}

The photon beamline has been designed to deliver extreme ultra-violet (XUV) photons in the energy range 250--3000~eV with photon fluxes exceeding $10^{12}$~s$^{-1}$ at resolving powers of at least 10000 \cite{Viefhaus2013}. The present measurements were taken in the commissioning phase of the beamline. In particular, the grating of the monochromator had not yet been aligned properly. This resulted in a lower photon flux and a lower achievable photon energy resolving power. During the present measurements the photon energy spread was $\Delta E \approx 3$~eV at photon energies around 700~eV. Due to the moderate resolving power the photon flux at this energy was still $5\times10^{12}$~s$^{-1}$. It was  monitored with a calibrated silicon photodiode  (labeled PD in figure~\ref{fig:PIPE}) located at the end of the photon beamline. The calibration had been performed off-site by the German national Metrology Institute Physikalisch-Technische Bundesanstalt (PTB). The uncertainty of the calibration is less than 1.5\% for photon energies in the range 250--3600~eV.

\subsection{Ion beamline}

In the PIPE setup (\fref{fig:PIPE}) an ion beam interacts with the photon beam either in a merged-beams or a crossed-beams arrangement. The latter provides a small interaction volume, e.g., for time-of-flight detection of photo products including photoelectrons over virtually the entire $4\pi$ solid angle. It is equipped with a recoil-ion momentum spectrometer \cite{Ullrich2003a} which was not used for the present measurements and will be described in detail elsewhere. The merged-beams section provides a large interaction volume and, thus, the possibility to study small cross sections of photon-ion processes with high sensitivity making up for the diluteness of ionic targets.

\begin{table}
\caption{\label{tab:overview}Overview over the investigated combinations of primary ion charge states $q$ and product-ion charge states $r$, and the corresponding experimental conditions. Listed are the ion current $I_\mathrm{ion}$ in the merged-beams section, the ion velocity $v_\mathrm{ion}$, the ion flight time $t_\mathrm{ion}$ from the ion source to the center of the voltage-labeled interaction region ($\sim$9~m flight path), and the experimental photon energy range $[E_\mathrm{min},E_\mathrm{max}]$.}
  \begin{indented}
    \item[] \begin{tabular}{@{}cccccrr}
    \br
     $q$ &  $I_\mathrm{ion}$ & $v_\mathrm{ion}$     & $t_\mathrm{ion}$ & $r$ & $E_\mathrm{min}$ & $E_\mathrm{max}$ \\
         &  (nA)             & (m~$\mu$s$^{-1}$) &    ($\mu$s) &     &   (eV)      & (eV)  \\
    \mr
  1 & 16.0 & 0.094 & 96 &4 &  660 & \phantom{0}750  \\
    & & && 5 &  500 & 1100 \\
    & & && 6 &  600 & 1100 \\
& & & & &\\
  2 & \phantom{1}2.0& 0.132 & 68 &5 &  660 & \phantom{1}750   \\
    & & && 6 &  660 & \phantom{1}750  \\
    & & && 7 &  660 & \phantom{1}750  \\
& & & & &  \\
  3 & \phantom{1}3.0 & 0.162 & 55 & 6 &  650 & \phantom{1}800  \\
    & & && 7 &  640 & \phantom{1}750  \\
    & & && 8 &  650 & \phantom{1}750  \\
& & & &  \\
  4 & \phantom{1}4.0 & 0.187 & 48 & 6 &  650 & 1200   \\
    & & && 7 &  650 & \phantom{1}750    \\
    & & && 8 &  650 & \phantom{1}770    \\
    & & && 9 &  660 & \phantom{1}770    \\
& & & &   & \\
  5 & \phantom{1}6.0 & 0.210 & 43 & 7 &  650 & 1000    \\
    & & & & 8 &  660 & \phantom{1}760    \\
    & & & &9 &  670 & \phantom{1}760    \\
    \br
  \end{tabular}
  \end{indented}
\end{table}

For the present measurements, Xe$^{q+}$ ($q=1-5$) ion beams were produced in a compact permanent-magnet electron cyclotron-resonance (ECR) ion source \cite{Broetz2001}. It had been installed on the ion-source platform (\fref{fig:PIPE}) and was operated on a high voltage $U_\mathrm{acc}$ with respect to the grounded ion beamline.  Thus, after extraction from the ECR ion source, ions with charge $qe$ (with $e$ denoting the elementary charge) and mass $m_\mathrm{ion}$ were accelerated to a laboratory energy $E_\mathrm{ion} = qeU_\mathrm{acc} = \frac{1}{2} m_\mathrm{ion}v_\mathrm{ion}^2$. The presently installed high-voltage power supply limits $U_\mathrm{acc}$ to a maximum value of  12.5~kV. For the present experiments $U_\mathrm{acc} = 6$~kV was used resulting in the tabulated (\tref{tab:overview}) ion velocities $v_\mathrm{ion}$.

\begin{figure}[t]
\begin{indented}
\item[]\includegraphics[width=0.5\textwidth]{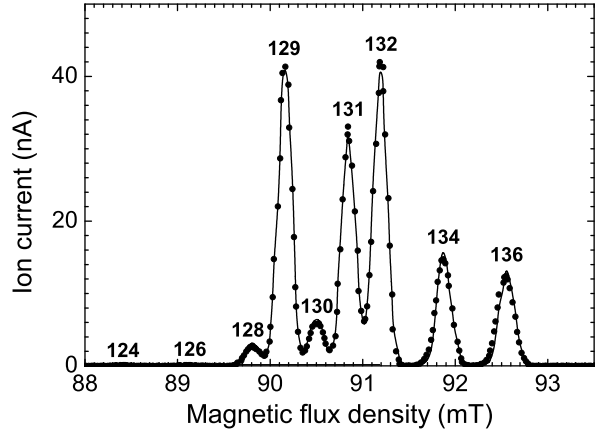}
\end{indented}
\caption{\label{fig:Xe2mass} Mass spectrum of a Xe$^{2+}$ ion beam  measured by scanning the analyzer magnet coil current and simultaneously  recording the transmitted ion current in a Faraday cup which was moved into the ion beam at about the downbeam focus of the analyzing magnet. Peaks are labeled by atomic mass number $A$. The measured
Xe$^{2+}$ mass distribution (symbols) corresponds to the natural abundance pattern and a mass resolving power $m/\Delta m \approx 250$ (full line).}
\end{figure}

A double focusing dipole magnet with $90^\circ$ bending angle, 1~m bending radius, and 2~m focal length is used for the selection of ions according to their mass-over-charge ratio. This \lq\lq analyzing magnet\rq\rq\ (\fref{fig:PIPE}) has been designed to deflect 1.5-keV-ions with masses of up to 100\,000~u for photoionization experiments. For example, singly charged mass-selected Fe cluster ions with more than 1500 atoms can be investigated. For the relatively light xenon ions in the present experiment isotopic mass resolution is routinely achieved (\fref{fig:Xe2mass}). Higher mass resolving powers can be realized at the expense of beam intensity by closing beam-collimating slits that are located at the focal points in front of and behind the analyzing magnet. In the present experiment ions with charge states $q=1-5$ and mass number $A=132$ were selected for further transport to the merged-beams interaction region.

The ions were directed onto the photon-beam axis by using an electrostatic spherical deflector. The length of the straight merging path is about 1.7~m. The ion beam was centered onto the counter-propagating photon beam by applying appropriate voltages to the electrostatic ion optical lenses and steerers that are placed along the ion beamline according to results from ion-optical calculations. In addition, slits can be closed onto the photon beam in order to force the thus collimated ion beam to the correct position.  A well defined interaction volume of length $L=49.5$~cm is provided by a drift tube that is coaxial with the merged photon and ion beams. When the drift tube is on an electric potential, $U_\mathrm{dt}$, photoions that are created inside the drift tube acquire a kinetic energy different from the kinetic energy of ions, that are created elsewhere along the merging path.

\begin{figure}[t]
\begin{indented}
\item[]\includegraphics[width=0.5\textwidth]{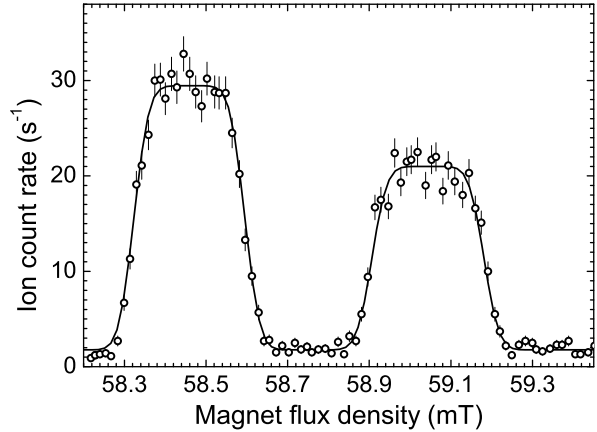}
\end{indented}
\caption{\label{fig:demergscan}  Demerging-magnet scan of Xe$^{8+}$ ionization products from a Xe$^{5+}$ primary ion beam interacting with 750-eV-photons. Xe$^{8+}$ ions that originate from the voltage-labeled interaction region ($U_\mathrm{dt} = -200$~V) contribute to the left peak and ions that are produced outside this region contribute to the right peak.  The full line is a fit to the measured  data (symbols) of two rectangular profiles convoluted with a gaussian. A gaussian full width at half-maximum of $0.029\pm0.001$~mT corresponding to a mass resolving power of $\sim$2000 has been obtained from the fit. }
\end{figure}

For the demerging of ions and photons a second double-focussing dipole magnet with a bending radius of 0.625~m is used. The demerging magnetic field deflects the ions from the photon beam axis and separates product ions (photoions) from primary ions. The primary ions are collected in a Faraday cup which is located inside the demerging-magnet vacuum chamber. With an appropriately chosen demerging magnetic flux density, product ions with a given energy and a given mass-over-charge ratio are deflected by $90^\circ$ and directed onto a single-particle detector. Before the ions reach the detector they are directed out of the collision plane by an electrostatic $180^\circ$ spherical deflector in order to suppress background from stray particles. The detector \cite{Rinn1982} consists of a metal converter plate where the impinging ions generate secondary electrons which by an electric field are accelerated into a channel electron multiplier (CEM).  Its detection efficiency for keV atomic ions is $\eta = 97\pm 3\%$ \cite{Rinn1982}. Dead time effects were negligible since the signal count rates were less than 200~s$^{-1}$ which is much less than the maximum count rate of several hundred kHz that can be processed. Under the present vacuum conditions the dark count rate was 0.036~s$^{-1}$.

\Fref{fig:demergscan} shows measured product-ion count rates resulting from triple photoionization of Xe$^{5+}$ primary ions as function of the demerging magnetic field. While scanning the magnetic field strength, Xe$^{8+}$ product ions originating from different parts of the photon-ion merging section were directed onto the detector. The low-field peak was caused by ions originating from the voltage-labeled interaction region. Since the drift-tube potential had been chosen to be negative ($U_\mathrm{dt} = -200$~V), these ions were somewhat slower than ions that originated from outside the voltage-labeled zone. Correspondingly, the latter have been detected at higher magnetic field values. The individual product ion peaks exhibit flat tops. This peak shape clearly shows that the entire product beam from the voltage-labeled interaction region reached the detector at the appropriate demerging magnetic setting.

\subsection{Cross-section determination}

\begin{figure}[t]
\begin{indented}
\item[]\includegraphics[width=0.5\textwidth]{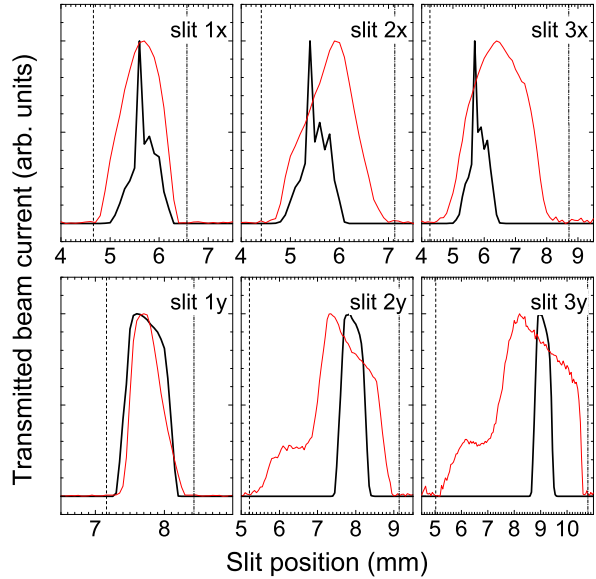}
\end{indented}
\caption{\label{fig:FF} (color online) Horizontal (upper panels) and vertical (lower panels) Xe$^{5+}$ ion-beam profiles (thin red lines) and photon-beam profiles (thick black lines) from slit-scan measurements. The slits are located (see also \fref{fig:FFL}) at $z_1 = -21.65$~cm (slits 1x and 1y), $z_2=0$~cm (slits 2x and 2y), and $z_3 = 21.65$~cm (slits 3x and 3y). The slit width is 5~$\mu$m for all slits. The dashed and dash-dotted vertical lines mark the integration ranges. Data points outside the integration range are used for background determination. The zero positions of the slits are arbitrary.}
\end{figure}

The photoionization cross section $\sigma$ is readily determined by normalizing the measured detector count rate $R$ on photon flux $\phi_\mathrm{ph}$, the number of ions in the interaction region, beam overlap and detection efficiency, i.e.,
\begin{equation}\label{eq:sigma}
 \sigma = R \frac{q\,e\,v_\mathrm{ion}}{\eta\,I_\mathrm{ion}\,\phi_\mathrm{ph}\,\mathcal{F}_L}
\end{equation}
where $I_\mathrm{ion}$ denotes the electrical ion current and where the beam overlap factor (form factor) for the length $L$ of the beam overlap region is defined as
\begin{equation}\label{eq:FF1}
    \mathcal{F}_L = \int_{-L/2}^{L/2}F(z)dz.
\end{equation}
At three positions $z_i$ ($i=1,2,3$) along the merging path the transverse beam-overlap factors
\begin{equation}\label{eq:FF2}
    F(z_i)=\frac{\int\!\!\int i_\mathrm{ion}(x,y)i_\mathrm{ph}(x,y)dxdy}{\int\!\!\int i_\mathrm{ion}(x,y)dxdy \int\!\!\int i_\mathrm{ph}(x,y)dxdy}
\end{equation}
can be determined by using scanning slits. At each position a set of two slits is available for scanning the beam profiles in horizontal and vertical direction, i.e., in $x$ and $y$ direction, respectively. \Fref{fig:FF} shows measured ion beam and photon beam profiles. Both beams overlapped well.

\begin{figure}
\begin{indented}
\item[]\includegraphics[width=0.5\textwidth]{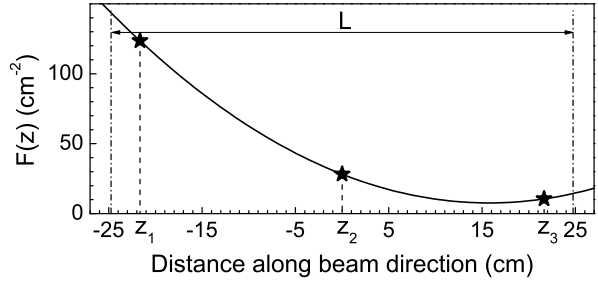}
\end{indented}
\caption{\label{fig:FFL} Determination of the beam overlap factor $\mathcal{F}_L$ with $L=49.5$~cm denoting the length of the voltage labeled interaction region.  The three form factors $F(z_i)$ (equation \ref{eq:FF2}) that were derived from the measured profiles in \fref{fig:FF} are plotted as symbols. The $z$-dependent form factor is interpolated by second order polynomial which is shown as thick full line. The integration along the $Z$-axis results in  $\mathcal{F}_L = 2233$~cm$^{-1}$. The integration range is marked by the two vertical dash-dotted lines at $z = \pm L/2$.}
\end{figure}

For the evaluation of the overlap factor $\mathcal{F}_L$, the integrals are approximated by sums and the usual approximation \cite{West2001a,Covington2002} is made that the ion current density and the photon flux density factorize such that $i_\mathrm{ion}(x,y)=i^m_\mathrm{ion}(x)i^m_\mathrm{ion}(y)$ and $i_\mathrm{ph}(x,y)=i^m_\mathrm{ph}(x)i^m_\mathrm{ph}(y)$. Measured current line densities $i^m_\mathrm{ion}(x)$, $i^m_\mathrm{ion}(y)$, $i^m_\mathrm{ph}(x)$, and $i^m_\mathrm{ph}(y)$ are displayed in \fref{fig:FF}. The remaining integration along the beam direction $z$ in equation \ref{eq:FF1} is carried out by interpolating the three measured form factors $F(z_i)$ with a second order polynomial (\fref{fig:FFL}).

It should be noted, that the overlap factor in the present setup does not depend on the photon energy, since the photon beam transport was designed such that the photon beam is always at the same spatial position independent of photon energy. This has been verified by repeatedly measuring beam profiles at different photon energies. The measured overlap factors were constant within 2\% in agreement with the findings of Lyon \etal \cite{Lyon1986} at a different setup. These measurements also showed that the ion beam remained at its position for at least several hours. This was due to its rather tight collimation and to the long-term stability of the supplies for the ion optical elements. As a further systematic check, overlap factors were measured for different settings of the ion-beam collimating slits. Within the experimental uncertainties, the product $I_\mathrm{ion}\mathcal{F}_L$ (cf.~Eq.~\ref{eq:sigma}) turned out to be independent of the collimator settings. Because of these highly stable conditions the measured cross sections could readily be put on an absolute scale.

The systematic uncertainty of the absolute cross section scale amounts to 15\% at 90\% confidence level. It is dominated by the 14\% uncertainty \cite{Covington2002} of the beam overlap factor.
Other major sources of error are the ion detection efficiency of the product ions (3\%), the collection efficiency of the primary ions (2\%), the ion current measurement (2\%), the photocurrent measurement (2\%), and the photodiode calibration (1.5\%).

\subsection{Energy calibration}\label{sec:cal}

An accurate assignment of resonance energies in the soft x-ray region is not straightforward. This is mainly due to the lack of suitable, sufficiently accurate calibration standards in this photon energy range. Nominal photon energies $h\nu_n$ on the high-resolution energy scale as calculated from the monochromator settings have uncertainties of less than 0.1\%. This was checked by measuring the energies of photoelectrons from photoemission of noble gases. A more accurate calibration was achieved by using C$^+$ and Ne$^+$ photoionization resonances at 288.40~eV and 848.66~eV, respectively \cite{Mueller2014}. This calibration resulted in an energy shift of -0.232~eV at 700~eV nominal energy. The remaining uncertainty of the calibrated energies $h\nu_c$ is mainly due to the misalignment of the monochromator grating which resulted in a slightly asymmetric instrument function of the photon beamline. Since this function is unknown, there is an associated uncertainty of resonance energies. A conservative estimate for the uncertainty of the present photon energy scale is 0.3~eV.

Finally, the Doppler shift of photon energies as seen by the moving ions has to be taken into account. To this end, the photon energy $h\nu$ in the rest frame of the primary ions is calculated from the calibrated photon energy $h\nu_c$ as
\begin{equation}\label{eq:Doppler}
h\nu = h\nu_c\sqrt{\frac{1+\beta}{1-\beta}}
\end{equation}
where $\beta = \sqrt{{2qe(U_\mathrm{acc}-U_\mathrm{dt})}/{(m_\mathrm{ion}c^2})}$
is the velocity of the primary ions in the voltage labeled interaction region in units of the speed of light $c$. Under the present experimental conditions the Doppler shift for Xe$^{1+}$ (Xe$^{5+}$) amounts to $\sim$0.22~eV (0.50~eV) at a photon energy of 700~eV.

\subsection{Metastable ions in the ion beam}

\begin{table}
\caption{\label{tab:exci}Calculated  excitation energies $E_\mathrm{DF}$ and life times $\tau_{DF}$ of levels from the Xe$^{q+}$ ground configurations for $q=1-5$. The calculated Dirac-Fock energies $E_\mathrm{DF}$ are compared with values $E_\mathrm{NIST}$ from the NIST atomic spectra database \cite{Kramida2013}.}
  \begin{indented}
    \item[]\begin{tabular}{@{}clllc}
    \br
     $q$ & level & {$E_\mathrm{NIST}$}  & {$E_\mathrm{DF}$} & $\tau_{DF}$ (ms)  \\
         & & {(eV)} & {(eV)} & {(s)}  \\
    \mr
  1 & $5s^2\,5p^5\;^2P_{3/2}$    & 0.0     &  0.0     &     $\infty$ \\
  1 & $5s^2\,5p^5\;^2P_{1/2}$    & 1.306   &  1.336   &     \phantom{4}44.3\\
    &  & & &\\
  2 & $5s^2\,5p^4\;^3P_{2}$      & 0.0     &  0.0     &     $\infty$ \\
  2 & $5s^2\,5p^4\;^3P_{0}$      & 1.008   &  1.107   &     $\infty$ \\
  2 & $5s^2\,5p^4\;^3P_{1}$      & 1.214   &  1.197   &     \phantom{4}54.4\\
  2 & $5s^2\,5p^4\;^1D_{2}$      & 2.120   &  2.441   &     \phantom{4}38.0\\
  2 & $5s^2\,5p^4\;^1S_{0}$      & 4.476   &  5.090   &     \phantom{43}4.1\\
    &  & & &\\
  3 & $5s^2\,5p^3\;^4S_{3/2}$    & 0.0     &  0.0     &     $\infty$ \\
  3 & $5s^2\,5p^3\;^2D_{3/2}$    & 1.645   &  2.150   &     \phantom{4}44.7\\
  3 & $5s^2\,5p^3\;^2D_{5/2}$    & 2.171   &  2.691   &     416\phantom{.0}\\
  3 & $5s^2\,5p^3\;^2P_{1/2}$    & 3.476   &  4.256   &     \phantom{4}14.4\\
  3 & $5s^2\,5p^3\;^2P_{3/2}$    & 4.420   &  5.139   &     \phantom{43}5.8\\
    &  & & &\\
  4 & $5s^2\,5p^2\;^3P_{0}$      & 0.0     & 0.0      &     $\infty$ \\
  4 & $5s^2\,5p^2\;^3P_{1}$      & 1.152   & 1.080    &     \phantom{4}91.6\\
  4 & $5s^2\,5p^2\;^3P_{2}$      & 1.751   & 1.810    &     434\phantom{.0}\\
  4 & $5s^2\,5p^2\;^1D_{2}$      & 3.522   & 3.798    &     \phantom{4}15.6\\
  4 & $5s^2\,5p^2\;^1S_{0}$      & 5.514   & 6.278    &     \phantom{43}3.3\\
    &  & & &\\
  5 & $5s^2\,5p\;^2P_{1/2}$      & 0.0     & 0.0      &     $\infty$ \\
  5 & $5s^2\,5p\;^2P_{3/2}$      & 1.934   & 1.952    &     28.6\\
    \br
  \end{tabular}
  \end{indented}
\end{table}

Since the ions are extracted from a hot plasma inside the ion source they are partially in excited levels. Most of these levels are short lived and decay rapidly to lower lying levels. However, few levels have considerably longer life times. We have calculated life times of the excited levels of the Xe$^{q+}$ ground configurations (table~\ref{tab:exci}) using the \textsc{Grasp} \cite{Joensson2013} Dirac-Fock atomic structure code. The calculated life times are orders of magnitude larger than the flight times from the ion source to the interaction region (table~\ref{tab:overview}). Thus, one has to be aware that the photoionization target has been a mixture of ions in the ground level and of ions in excited levels. As usual, the fractions of excited ions are unknown. Correspondingly, any comparison between experimental and theoretical results has to be discussed with due care. It should, however, be noted that the excited level energies  (table~\ref{tab:exci}) are very much lower than the presently used photon energies. Therefore, we do not expect a strong influence of the excitation state on the measured cross sections, in particular, in photon energy ranges where there are no ionization thresholds or photoionization resonances.

\section{Results and discussion}\label{sec:res}

\begin{figure}
\begin{indented}
\item[]\includegraphics[width=0.5\textwidth]{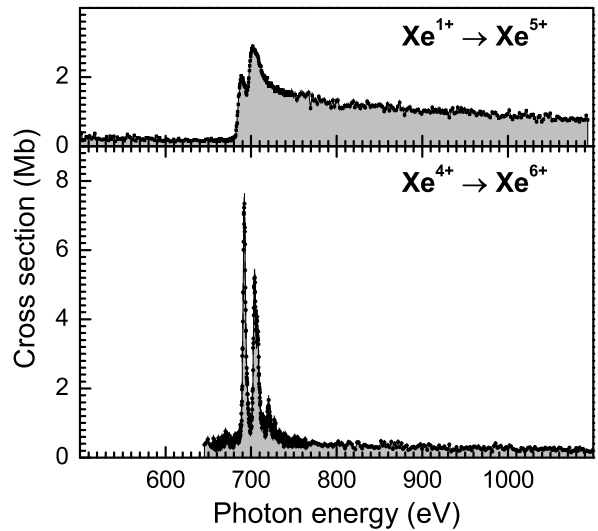}
\end{indented}
\caption{\label{fig:Xe14} Measured absolute cross sections for fourfold photoionization of Xe$^{+}$ and double photoionization of Xe$^{4+}$ions.}
\end{figure}

\begin{figure}[t]
\begin{indented}
\item[]\includegraphics[width=0.5\textwidth]{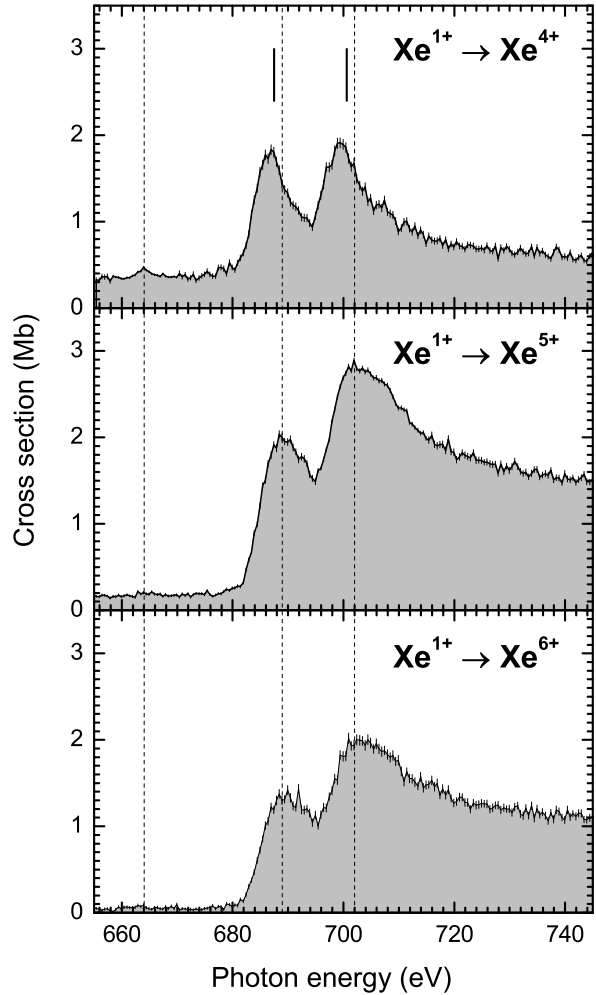}
\end{indented}
\caption{\label{fig:Xe1to456} Measured absolute cross sections for multiple photoionization of Xe$^{+}$ ions (shaded curves). The full vertical lines in the upper panel denote the threshold energies for $3d_{5/2}$ and $3d_{3/2}$ ionization of Xe$^+$ from table~\ref{tab:thres}. The dashed vertical lines are drawn to guide the eye.}
\end{figure}

\begin{figure}
\begin{indented}
\item[]\includegraphics[width=0.5\textwidth]{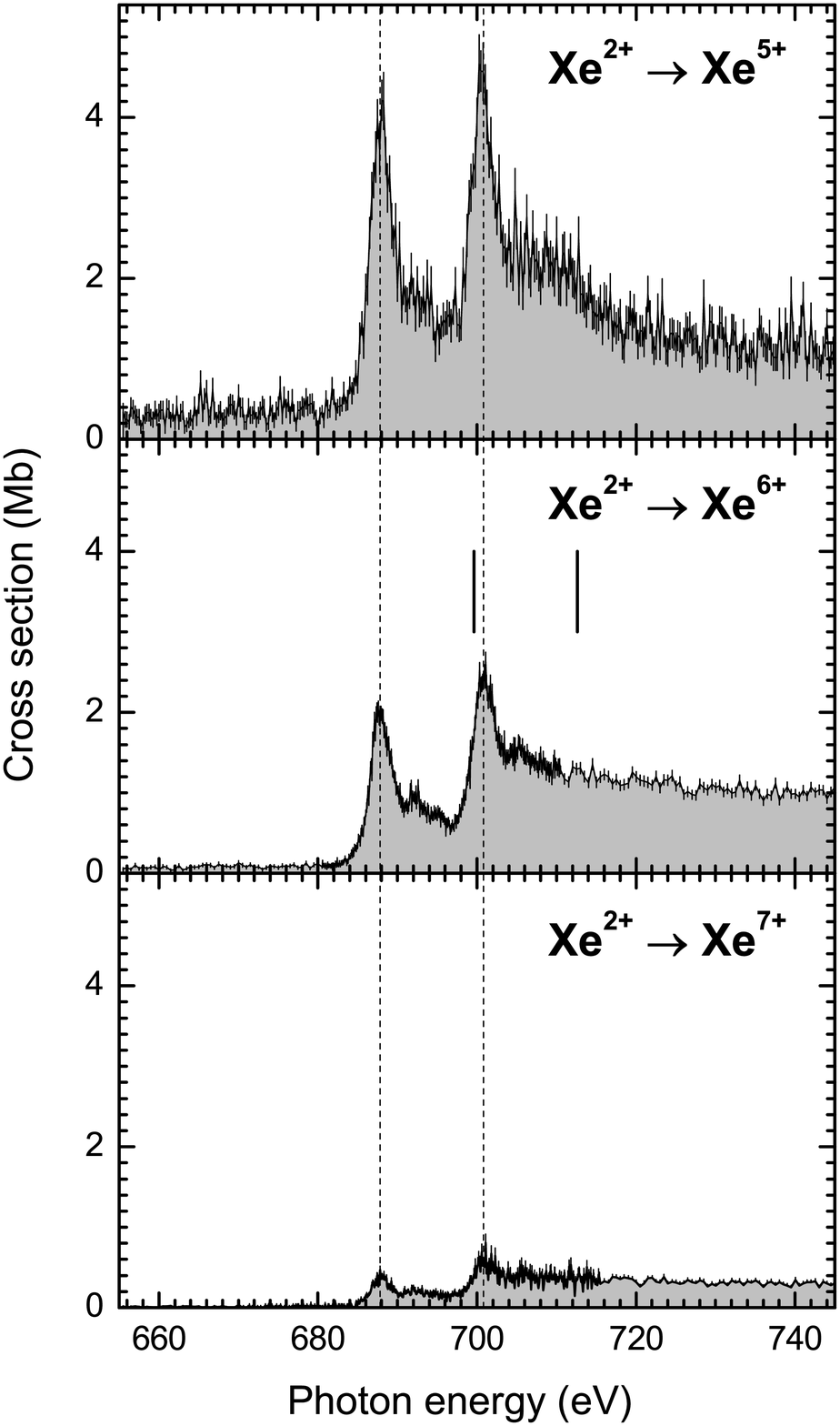}
\end{indented}
\caption{\label{fig:Xe2to567} Measured absolute cross sections for multiple photoionization of Xe$^{2+}$ ions. The full vertical lines in the center panel denote the threshold energies for $3d_{5/2}$ and $3d_{3/2}$ ionization of Xe$^{2+}$ from table~\ref{tab:thres}. The dashed vertical lines are drawn to guide the eye.}
\end{figure}

\begin{figure}
\begin{indented}
\item[]\includegraphics[width=0.5\textwidth]{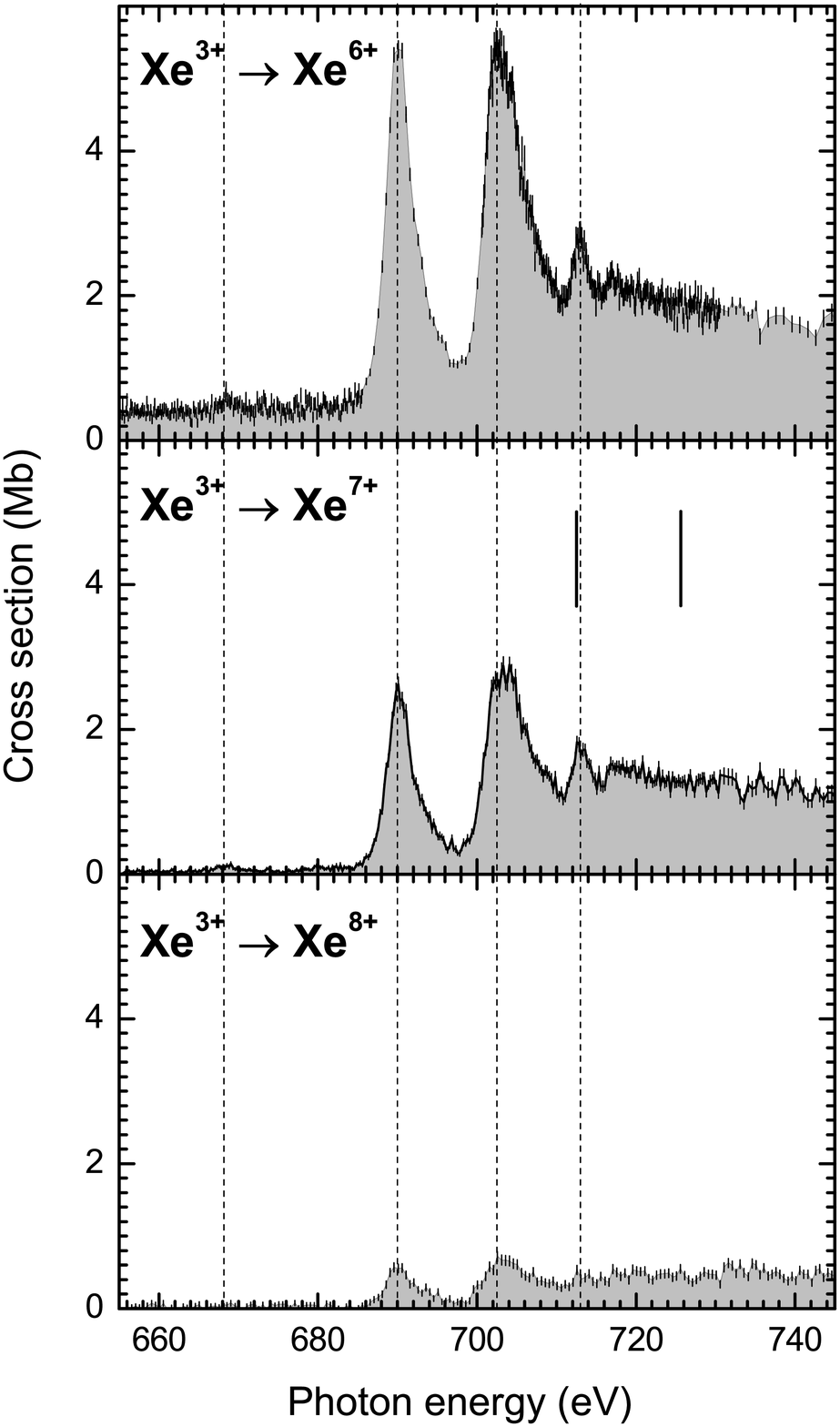}
\end{indented}
\caption{\label{fig:Xe3to678} Measured absolute cross sections for multiple photoionization of Xe$^{3+}$ ions. The full vertical lines in the center panel denote the threshold energies for $3d_{5/2}$ and $3d_{3/2}$ ionization of Xe$^{3+}$ from table~\ref{tab:thres}. The dashed vertical lines are drawn to guide the eye.}
\end{figure}

\begin{figure}
\begin{indented}
\item[]\includegraphics[width=0.5\textwidth]{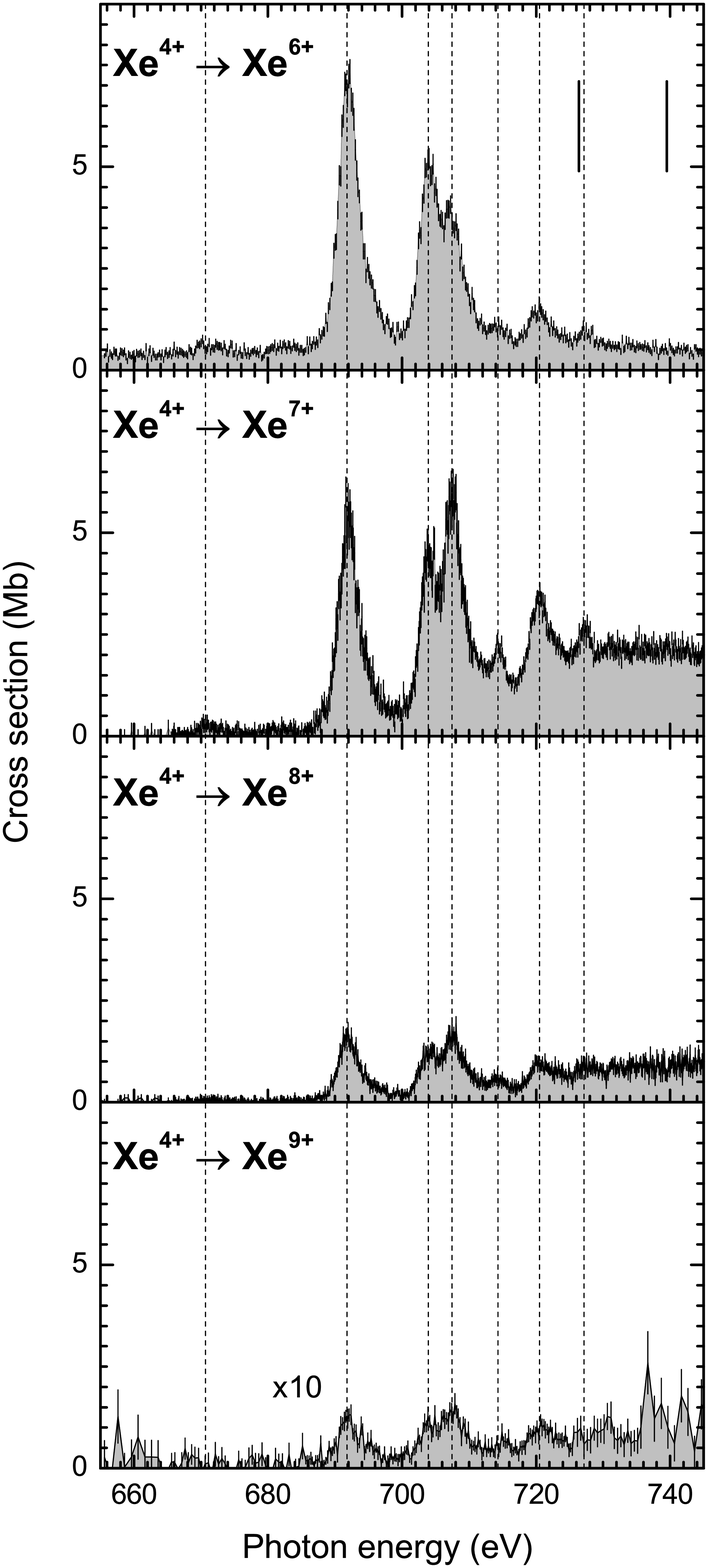}
\end{indented}
\caption{\label{fig:Xe4to6789} Measured absolute cross sections for multiple photoionization of Xe$^{4+}$ ions. The full vertical lines in the top panel denote the threshold energies for $3d_{5/2}$ and $3d_{3/2}$ ionization of Xe$^{4+}$ from table~\ref{tab:thres}. The dashed vertical lines are drawn to guide the eye.}
\end{figure}

\begin{figure}
\begin{indented}
\item[]\includegraphics[width=0.5\textwidth]{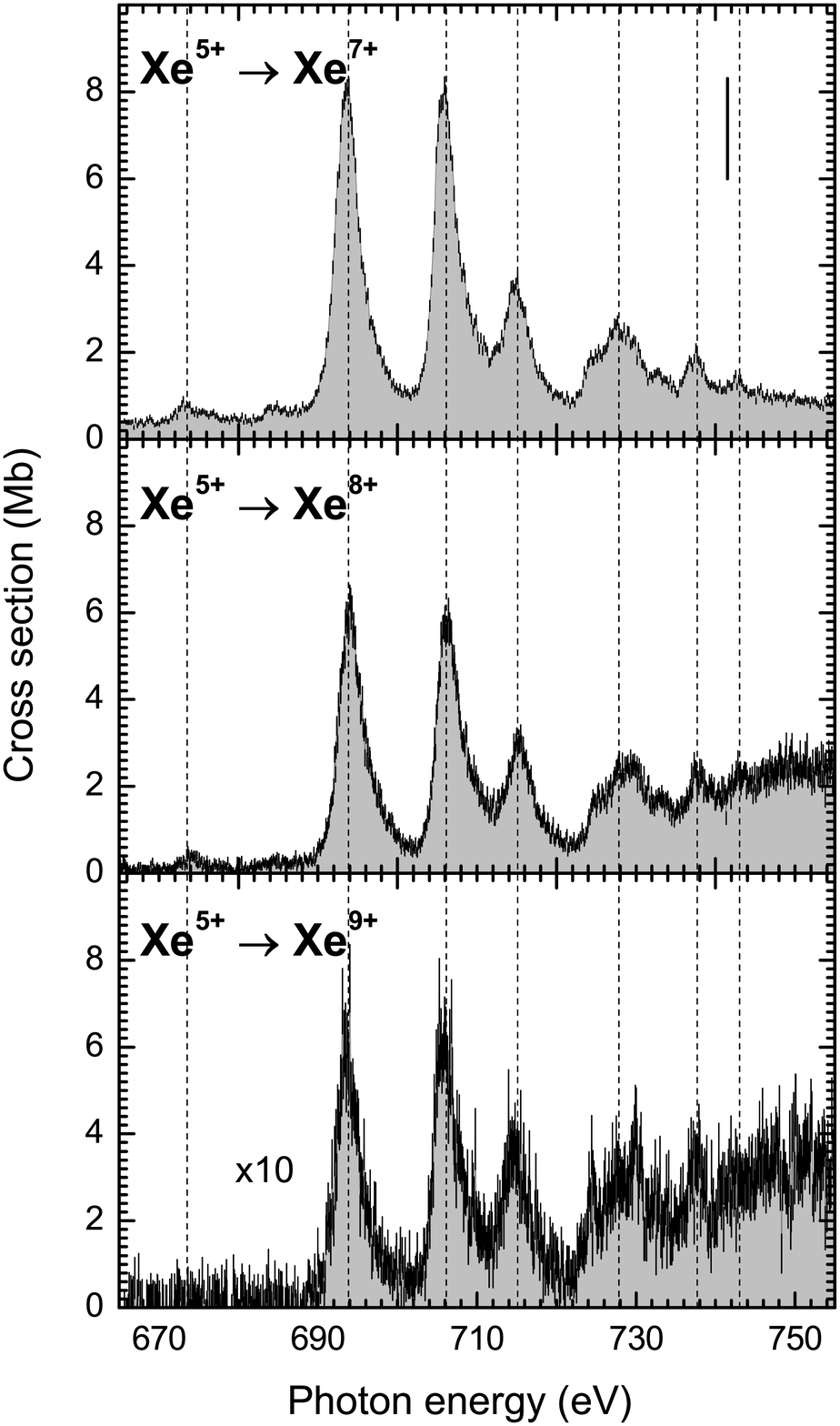}
\end{indented}
\caption{\label{fig:Xe5to789} Measured absolute cross sections for multiple photoionization of Xe$^{5+}$ ions. The full vertical line in the top panel denotes the threshold energies for $3d_{5/2}$ ionization of Xe$^{5+}$ from table~\ref{tab:thres}. The dashed vertical lines are drawn to guide the eye.}
\end{figure}

Figure \ref{fig:Xe14} shows cross sections for fourfold ionization of Xe$^+$ and double ionization of Xe$^{4+}$ over rather extended photon energy ranges of 500--1100~eV and 650--1100~eV, respectively. The two spectra are strikingly different, in particular, at energies around 700~eV. While the Xe$^+$ spectrum exhibits a double-step structure with an extended high energy tail, the Xe$^{4+}$ spectrum shows distinct resonance features which are mainly associated with $3d\to nf$ excitations. This striking difference is related to the well-known collapse of the $nf$ wave functions \cite{Cheng1983} which occurs when the charge of the primary ion is increased. The progression towards spectra containing increasingly more distinct resonance features with increasing primary ion charge state is displayed in detail in figures~\ref{fig:Xe1to456}--\ref{fig:Xe5to789}. Each measured curve represents an absolute cross section for multiple photoionization, i.e., for a reaction of the type
\begin{equation}\label{eq:reaction}
    h\nu + \mathrm{Xe}^{q+} \to \mathrm{Xe}^{r+} + (q-r)e^-,
\end{equation}
where $q$ and $r$ denote the charge states of the primary ion Xe$^{q+}$ and the detected product ion Xe$^{r+}$, respectively. In figures~\ref{fig:Xe1to456}--\ref{fig:Xe5to789} each spectrum carries a corresponding label \lq Xe$^{q+}$$\to$Xe$^{r+}$\rq. For each primary ion only those product ion charge states were considered which yielded a count rate larger than $\sim$1 Hz at the cross section maximum. This condition constrained the measured product ion charge states $r$ to the ones given in table \ref{tab:overview}.

\begin{table}
\caption{\label{tab:thres}Calculated (see text) threshold energies (configuration averages, in eV) for $3d_{5/2}$ and $3d_{3/2}$ ionization of Xe$^{q+}$ primary ions with $0\leq q \leq 5$. The values for $q=0$ are in good agreement with the spectroscopic values of 676.70~eV and 689.35~eV \cite{Gelius1974,Svensson1976}.}
  \begin{indented}
    \item[]\begin{tabular}{@{}lll}
    \br
    $q$ &  $3d_{5/2}$ & $3d_{3/2}$ \\
    \mr
    0 & 676.7 & 689.7 \\
    1 & 687.5 & 700.6 \\
    2 & 699.6 &	712.6 \\
    3 & 712.5 &	725.6 \\
    4 & 726.4 &	739.5 \\
    5 & 741.5 &	754.6 \\
    \br
  \end{tabular}
  \end{indented}
\end{table}

\begin{figure}
\begin{indented}
\item[]\includegraphics[width=0.5\textwidth]{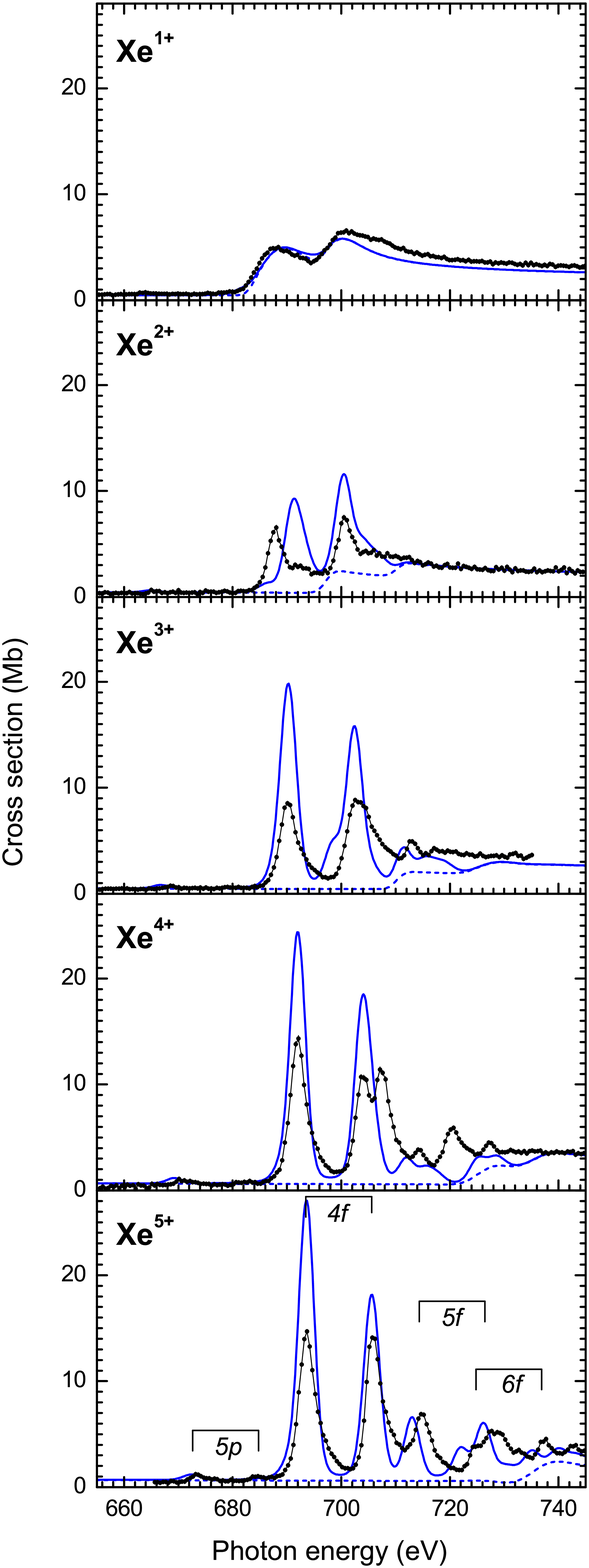}
\end{indented}
\caption{\label{fig:Sum} (color online) Experimental sum spectra (symbols) and results of the present Dirac-Fock calculations (blue full lines). The dashed lines represent the cross sections for direct ionization only. The theoretical spectra were shifted by -3.5 eV (Xe$^+$) and -2.3~eV (Xe$^{2+}$--Xe$^{5+}$) in order to align them to the experimental resonance structure. The theoretical spectra have been convoluted with a normalized gaussian with 3.0~eV full-width-at-half-maximum to mimic the influence of the experimental photon energy spread.}
\end{figure}

In order to obtain insight into the measured spectra we performed fully relativistic  Dirac-Fock atomic structure calculations using the latest versions of the \textsc{Grasp} \cite{Joensson2013} and \textsc{Ratip} \cite{Fritzsche2012a} codes. Initial wave functions for the Dirac-Fock calculations were obtained from the Hartree-Fock  program \textsc{HF} of Froese Fischer \cite{FroeseFischer1987}. The calculations were kept as simple as possible, i.e., all atomic state functions were based on just one nonrelativistic configuration. Thus, the present theoretical results are not of utmost precision. Nevertheless they are well suited for the discussion of the main features in the experimental spectra. The natural line widths of the $3d$ excitation resonances are mainly determined by the lifetime of the $3d$ hole, which is quickly filled by Auger transitions. From our atomic-structure calculations we found a width of 0.6~eV for virtually all resonances independent of the ion charge state. This width is within the range of values reported for neutral xenon \cite{Arp1999}. Results of our calculations are presented in tables~\ref{tab:exci} and \ref{tab:thres} and in figure~\ref{fig:Sum}.

For Xe$^+$ the dominant product ion charge states are ${4+}$,${5+}$, and ${6+}$ (figure~\ref{fig:Xe1to456}). The same charge states have been found to be dominant for $3d$ ionization of neutral xenon \cite{Saito1992}. Also the shapes of the present cross sections for Xe$^{+}$ strongly resemble measured absorption and ionization cross sections for neutral xenon atoms \cite{Deslattes1968,Yagci1983,Sonntag1984,Saito1992,Arp1999,Kivimaeki2000}. In the energy range 670--720~eV, a broad double-peak structure is observed on top of a monotonically decreasing background cross section due to outer shell ionization. The energy separation of the two peaks corresponds to the $3d_{5/2}-3d_{3/2}$ fine-structure splitting of about 13~eV. This splitting is virtually the same for all presently investigated charge states (table \ref{tab:thres}). In analogy to $3d$ ionization of neutral Xe, the double peak structure in the Xe$^+$ cross sections is attributed mainly to $3d\to\varepsilon f$ shape resonances \cite{Manson1968} which are supported by a shallow centrifugal barrier of the atomic potential. A competing ionization mechanism is resonant excitation of the $3d$ electron to a higher $nf$ shell with principal quantum numbers $n\geq 4$ and subsequent autoionization. For neutral Xe this ionization pathway is largely suppressed because the $nf$ wave functions are localized in the outer potential well beyond the centrifugal barrier. Therefore, they do not have a significant spatial overlap with the $3d$ wave function. For Xe$^+$ the rise of the ionization cross section occurs just at the $3d_j$ ($j=3/2, 5/2)$ ionization thresholds. This is in contrast to neutral xenon where a \lq\lq delayed\rq\rq\ onset of the absorption cross section is observed at about 10~eV above the threshold \cite{Deslattes1968,Manson1968}.

The cross sections for the higher primary ion charge states exhibit narrower $3d_j\to n\ell$ excitation resonances in addition to the broader threshold features. The resonances appear below the corresponding $3d_j$ ionization thresholds. The number of individually resolved resonance features increases with increasing charge state of the primary ion (figures~\ref{fig:Xe2to567}--\ref{fig:Xe5to789}). While the positions of resonances of a given primary charge state are independent of the product charge state, the relative resonance strengths differ from one product channel to another. This is most noticeable in the double and triple ionization spectra of Xe$^{4+}$ (figure~\ref{fig:Xe4to6789}). Generally, the individual resonance strengths depend on the details of the various decay chains that lead from the primary $3d$ hole to the different product ions.

The slight differences between the cross sections for threefold and fourfold ionization of Xe$^+$ (figure~\ref{fig:Xe1to456}) may be explained similarly. The apparent shift of the $3d\to\varepsilon f$ resonance features in the  Xe$^+$$\to$Xe$^{4+}$ cross section to lower energies  may  be caused by unresolved resonances which are much suppressed in the Xe$^+$$\to$Xe$^{5+}$ and Xe$^+$$\to$Xe$^{6+}$ cross sections. In addition, the Xe$^+$$\to$Xe$^{4+}$ cross section exhibits a weak peak at 664~eV which may be associated with $3d_{5/2}\to5p$ excitation. This peak is not discernible in the other multiple ionization channels. It is noted that pathways after $3d$ photoionization of neutral Xe have been investigated in more detail by applying electron spectroscopy and coincidence techniques \cite{Jonauskas2003,Tamenori2002,Matsui2004,Partanen2005,Viefhaus2005,Kochur2009,Suzuki2011}.

For the comparison of the measured absolute cross sections with the calculated absorption cross sections we have summed all individually measured product channels for each primary ion. To a good approximation, these sum spectra represent the absolute experimental cross sections for photoabsorption. According to the findings of Saito and Suzuki \cite{Saito1992} for neutral Xe, the sum of the three dominant ionization channels comprises more than 90\% of total cross section for $3d$ photoionization. Therefore, we assume that the additional uncertainty of the experimental $3d$ photoabsorption cross sections due to the neglect of unmeasured product charge states is less than 10\%. 

Experimental and calculated spectra are compared in figure~\ref{fig:Sum}. It should be noted that experimental and theoretical cross sections are both absolute, i.e., there is no scaling of cross section involved in the comparison. In the energy ranges where direct $3d$ ionization is dominant, experimental and theoretical cross sections agree to within 20\% or better, i.e., within the experimental uncertainties (including the 10\% summation uncertainty). The largest discrepancies (at energies $> 730$~eV) of up to 20\% occur for Xe$^{1+}$ and Xe$^{3+}$. There is no significant discrepancy for the other charge states. The situation is different for energy ranges which are dominated by resonances. The theoretical spectra were shifted towards lower energies by up to 3.5~eV (figure~\ref{fig:Sum}) in order to achieve a gross alignment of experimental and theoretical resonance structures. Still most individual resonance structures do not match very well. There is disagreement with respect to resonance positions and resonance strengths. Apparently, our simple single-configuration approach neglects important correlation effects. In addition, there may have been sizeable fractions of metastable primary ions (table~\ref{tab:exci}) in the experiment with possibly non-negligible effects on the measured resonance structures.

Nevertheless, the calculations are very useful for assigning the measured resonances. Accordingly, the most dominant ionization resonances are associated with  $3d_{5/2}\to nf$ and $3d_{3/2}\to nf$ excitations ($n = 4, 5, \ldots$) with any pair of $3d\to nf$ excitations being separated by the $3d_{5/2}-3d_{3/2}$ fine-structure splitting of $\sim$13~eV. For Xe$^{5+}$, resonances up to $n=6$ have been resolved (bottom panel of figure \ref{fig:Sum}). At low energies, weak resonances associated with $3d\to5p$ excitations are clearly visible for Xe$^{4+}$ and Xe$^{5+}$. According to our calculations, $3d\to np$ excitations with $n>5$ are very weak for all primary ion charge states under consideration.

\section{Summary and conclusions}\label{sec:concl}

The new photon-ion merged-beams setup PIPE has successfully been used for measuring absolute cross sections for multiple $3d$ photoionization of ions from the xenon isonuclear sequence.
The measured cross sections exhibit progressively more resonance features as the primary ion charge state is increased from 1+ to 5+. This finding is explained by the collapse of the $f$ wave functions into the inner atomic potential well which deepens with increasing charge state. Absolute experimental cross sections for photoabsorption were derived by summation over all measured ionization channels. The present results are useful for modeling of the charge balance in astrophysical and technical plasmas.

The main experimental cross section features were reproduced by single-configuration Dirac-Fock calculations. Accordingly, the resonances are attributed to $3d \to nf$ excitations with subsequent autoionization. Competing $3d\to np$ excitations were found to play only a minor role. The theoretical cross sections for direct ionization agree with the experimentally derived absorption cross sections within the experimental uncertainties. However, the calculated resonance structures differ significantly from the experimental findings. This is attributed mainly to the neglect of correlation effects in the theoretical single-configuration approach.

With the present investigation we hope to stimulate more accurate theoretical work on inner-shell photoionization of complex ions. Experimental spectra taken at higher resolving powers are already available. Their detailed analysis is in progress and will be the topic of a forthcoming publication.

\ack

Technical and financial support by DESY is gratefully acknowledged. The construction and building of the PIPE setup has been made possible by substantial funding from the German ministry for education and research (BMBF) under contracts 05KS7RG1, 05KS7GU2, 05KS7KE1, 05KS7RF2, 05K10RG1, 05K10GUB, 05K10KEA, and 05K10RF2 within the \lq\lq Verbundforschung\rq\rq\ funding scheme.

\section*{References}


\end{document}